\documentstyle[epsfig]{aipproc}
\def\lapproxeq{\lower .7ex\hbox{$\;\stackrel{\textstyle
<}{\sim}\;$}}
\def\gapproxeq{\lower .7ex\hbox{$\;\stackrel{\textstyle
>}{\sim}\;$}}
\def\funp{{I\!\!P}}
\def\be{\begin{equation}}
\def\ee{\end{equation}}
\def\bea{\begin{eqnarray}}
\def\eea{\end{eqnarray}}
\begin{document}
\title{A unified BFKL/DGLAP description of Deep Inelastic
Scattering}
\author{J.\ Kwieci\'{n}ski$^{*,\dagger}$, A.D.\ Martin$^{\dagger}$ and A.M.\
Stasto$^{*,\dagger}$ }
\address{$^*$ H.\ Niewodnicza\'{n}ski Institute of Nuclear Physics,
Department of Theoretical Physics,\\
 ul.\ Radzikowskiego 152, 31-342
Krakow, Poland.\\
$^{\dagger}$ Department of Physics, University of Durham, Durham, DH1
3LE, UK. }
\maketitle
\begin{center}
(presented by R.G.\ Roberts) 
\end{center}
\begin{abstract}
We introduce a coupled pair of evolution equations for the
unintegrated gluon distribution and the sea quark distribution
which incorporate both the resummed leading $\ln (1/x)$ BFKL
contributions and the resummed leading $\ln (Q^2)$ DGLAP
contributions.  We solve these unified equations in the
perturbative QCD domain.  With only two physically motivated
parameters we obtain an excellent description of the HERA $F_2$
data.  
\end{abstract}
\section*{1.~Introduction} 
One of the striking features of the measurements of deep
inelastic scattering at HERA is the strong rise of the proton
structure function $F_2$ as $x$ decreases from $10^{-2}$ to below
$10^{-4}$ \cite{DATA}.  At first sight it appeared that the rise
was due to
the (BFKL) resummation of leading $\ln (1/x)$ contributions.  In
this approach the basic dynamical quantity at small $x$ is the
gluon distribution $f (x, k_T^2)$ {\it unintegrated} over its
transverse momentum $k_T$.  Observables are computed in terms of
$f$ via the $k_T$ factorization theorem.  For example
\be
F_2 \; = \; F_2^{\gamma g} \: \otimes \: f \quad {\rm where}
\quad
F_2^{\gamma g} \: = \: \sum_q \: e_q^2 \:
B_q
\label{eq:a1}
\ee
and where $\otimes$ denotes a convolution in transverse ($k_T$),
as well as
longitudinal, momentum, see Fig.~1(a).  $F_2^{\gamma g}$ is the
off-shell gluon
structure function which at lowest order is given by the \lq \lq
quark box and crossed-box" contributions, $\gamma g \rightarrow q
\overline{q}$.
On the other hand the conventional DGLAP approach is based on
collinear factorization
\be
{\partial F_2\over x\partial ln(Q^2/\Lambda^2)}\; = \: \sum_i \: e_i^2  
\; P_{qg} \overline{\otimes} g +\: \sum_i \: e_i^2   P_{qq}
\overline{\otimes} (q_i+\bar q_i)
\label{eq:a2}
\ee
where $\overline{\otimes}$ is simply a convolution over
longitudinal
momentum.  DGLAP evolution effectively sums up the leading $\ln
Q^2$ contributions, and is able to describe the $F_2$ data at the
smallest $x$ values observed (even for $Q^2 \sim 1 {\rm GeV}^2$)
with an appropriate choice of input parton distributions.  These
non-perturbative input shapes in $x$ mean that there is more
freedom in the pure DGLAP description than in the BFKL approach. 
In the latter approach, the shape in $x$ emerges, in principle,
from the perturbative $\ln (1/x)$ resummation.  In practice it is
not so clear cut since non-perturbative, and also subleading $\ln
(1/x)$, effects can modify the BFKL prediction. 
Here we should attempt to find a unified description of $F_2$
which incorporates both (DGLAP and BFKL) of these perturbative effects.\\
\begin{figure} 
\centerline{\epsfig{file=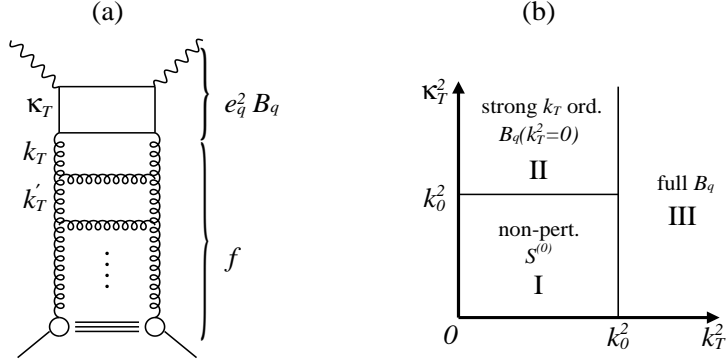,height=11cm,width=11cm}}
\vspace{-5cm}
\caption{(a) The diagrammatic representation of the $k_T$
factorization formula $F_2 \: = \: F_2^{\gamma g} \otimes f$.  At
lowest order $F_2^{\gamma g}$ is given by the quark box shown
(together with the crossed box) summed over all types of quark. 
(b) The different regions of integration used to evaluate the
sea quark distribution.  In regions II and III the sea is driven
by the gluon, $S_q = B_q \otimes f$.
}
\label{fig.1}
\end{figure}

\section*{2.~Unified BFKL/DGLAP formalism} 
We argue that the unintegrated gluon distribution $f (x, k_T^2)$
and the $k_T$ factorization theorem provides the natural
framework for describing observables at small $x$.  To determine
$f$ we arrange the BFKL equation so that we only need to solve it in
the
perturbative domain $k_T^2 > k_0^2$ \cite{KMS}.  We also include
the
residual DGLAP contributions.  To be precise we have
\bea
\label{eq:b1}
 f (x,k_T^2) & = & \frac{\alpha_S}{2 \pi} \: \int_x^1 d z P_{gg}(z)
\frac{x}{z} \; g (\frac{x}{z}, k_0^2) \: + \nonumber \\
\nonumber \\
& \hspace{-3cm} + & \hspace{-1.6cm} \frac{3 \alpha_S}{\pi} \; k_T^2 \int_x^1 \frac{dz}{z} \:
\int_{k_0^2} \frac{d
k_T^{\prime 2}}{k_T^{\prime 2}} \left[ \frac{f ({x \over z},
k_T^{\prime 2}) \: \theta \: ({k_T^2 \over z} - k_T^{\prime
2}) - f ({x \over z}, k_T^2)}{| k_T^{\prime 2} - k_T^2 |}
\; + \; \frac{f ({x \over z}, k_T^2)}{(4 k_T^{\prime 4} + 
k_T^4)^{\frac{1}{2}}} \right] \; \nonumber \\
\nonumber \\
& \hspace{-3cm} + & \hspace{-1.6cm} \frac{3 \alpha_S}{\pi} \:
\int_x^1 d z \: \left(\frac{P_{gg}(z)}{6} \: - \frac{1}{z} \right)
\int_{k_0^2}^{k_T^2} \frac{d k_T^{\prime 2}}{k_T^{\prime 2}} \; f
\left(\frac{x}{z}, k_T^{\prime 2} \right) \: + \:
\frac{\alpha_S}{2 \pi} \:
\int_x^1 d z P_{gq} \: \Sigma \left(\frac{x}{z}, k_T^2 \right)
\nonumber\\
\eea
where $-1/z$  is taken from DGLAP  because it is already included in
BFKL.  The input term comes from two sources: the $k_T^2 < k_0^2$
parts of BFKL and DGLAP terms.  We specify the input in terms
of a simple two parameter form
$g (x, k_0^2) \: = \: N (1 - x)^{\beta}$.
In addition to restricting the solution of the BFKL equation to
the perturbative region $k_T^2 > k_0^2$ and to including the
DGLAP terms, we have also introduced a $\theta$ function which
imposes the constraint $k_T^{\prime 2} < k_T^2 / z$ on the real
gluon emissions.  The origin of this constraint\footnote{A more
general treatment of the gluon ladder which incorporates both the
BFKL equation and DGLAP evolution is given by the CCFM equation
\cite{CCFM}, which is based on angular ordering of the gluon
emissions.  The angular ordered and kinematic constraints lead to
similar subleading $\ln (1/x)$ effects, but the kinematic
constraint overrides the angular ordered constraint, except when
$Q^2 < k_T^2$ in the large $x$ domain \cite{KMS1}.} is the
requirement
that the virtuality of the exchanged gluon is dominated by its
transverse momentum $| k^{\prime} |^2 \simeq k_T^{\prime 2}$. 
 We
take a running coupling $\alpha_S (k_T^2)$, which is supported by
the results of the next-to-leading order $\ln (1/x)$ analyses of
Fadin, Lipatov, Camici and Ciafaloni.
The final term in (\ref{eq:b1}) depends on the quark singlet
momentum distribution $\Sigma$.  At small $x$ the sea quark
components $S_q$ of $\Sigma$ dominate.  They are driven by the
gluon via the $g \rightarrow q \overline{q}$ transitions, that is
$S_q \; = \; B_q \otimes f$ 
where at lowest order $B_q$ is the box (and crossed box)
contribution indicated in Fig.~1(a).  Besides the $z$ and $k_T^2$
integrations symbolically denoted by $\otimes$ the box
contribution implicitly includes an integration over the
transverse momentum $\kappa_T$ of the exchanged quark.  The
evolution equation for $\Sigma$ may be written in the form
\be
\label{eq:b4}
\Sigma \; = \; S^{(0)} \: + \: \sum_q \: B_q (k_T^2 = 0) \:
\overline{\otimes} \: zg (z, k_0^2) \: + \: \sum_q B_q \otimes f
\: + \: P_{qq} \otimes S_q \: + \: V
\ee
where the first three terms on the right hand side are the \lq
\lq $B_q \otimes f$" contributions coming from
the regions I, II, III of the $k_T^2$ and $\kappa_T^2$
integrations that are shown in Fig.~1(b).  First, in the
non-perturbative domain, region I, the $u, d, s$ sea quark
contribution is parametrized in the form 
$S^{(0)} \; = \; C_{\funp} \: x^{-0.08} (1 - x)^8$
consistent with soft pomeron and counting rule expectations,
where $C_{\funp}$ is independent of $Q^2$.  The constant
$C_{\funp}$ is fixed in terms of the two parameters, $N$ and
$\beta$, by the momentum sum rule.  In region II
we apply the strong $k_T$ ordering approximation with $B_q
\approx B_q (k_T^2 = 0)$ so that the $k_T^2$ integration can be
carried out to give a contribution proportional to $g (x/z,
k_0^2)$.  Finally in region III we evaluate the full box
contribution; this gives the main contribution and is responsible
for the rise of $F_2$ with decreasing $x$.  The last two terms in
(\ref{eq:b4}) give the sea $\rightarrow$ sea evolution
contribution, and the valence contribution $V (x, Q^2)$ which is
taken directly from a recent parton set.  The charm quark component
of the sea is given totally by {\it perturbative} QCD, since for
$ k_T^2 < k_0^2$ the box $B (k_T^2 = 0)$ is finite as $\kappa_T^2
\rightarrow 0$ due to $m_c \neq 0$. \\

\section*{3.~Description of $F_2$ data and discussion}
We solve the coupled integral equations (\ref{eq:b1}) and
(\ref{eq:b4}) for the gluon $f$ and the quark singlet $\Sigma$ in
the perturbative domain, $k_T^2 > k_0^2$.  The only input is the
gluon $g (x, k_0^2)$. We take $k_0^2 = 1 {\rm
GeV}^2$.  We determine the values of the two input parameters by
fitting to the available data \cite{DATA} for $F_2$ with $x <
0.05$ and $Q^2 > 1.5 {\rm GeV}^2$.  The continuous curves in
Fig.~2 show the description of a sample of the data.  Overall the
fit is excellent; at least as good as that achieved in the recent
global analyses.  When we repeat the analysis with the kinematic
constraint omitted we see that the description (given by the
dashed curves in Fig.~2) is not so good and, moreover, the
extrapolation of the gluon to $x \approx 0.4$ no longer describes
the WA70 prompt photon data.
\begin{figure} 
\centerline{\epsfig{file=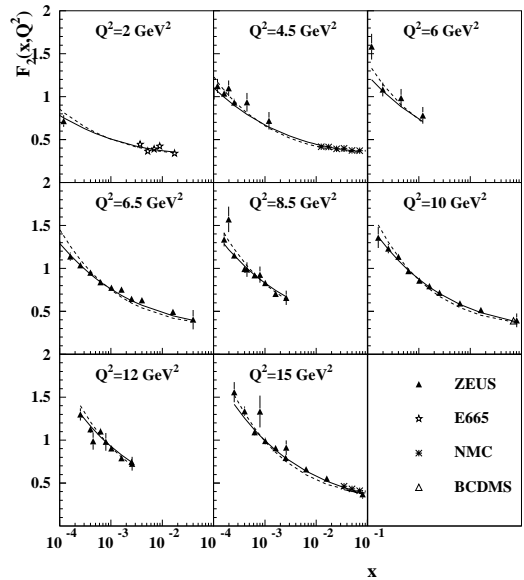,height=8cm,width=8cm}}
\caption{The two-parameter description of some of the $F_2$
data at small $x$ using $f (x,k_T^2)$ evaluated with (continuous
curves) and without (dashed curves) the kinematic constraint
$k_T^{\prime 2} < k_T^2 / z$.} 
\label{fig.2}
\end{figure}
How important are the $\ln (1/x)$ effects?  First we replace the
BFKL kernel in (\ref{eq:b1}) by the standard DGLAP splitting
function, $P_{gg}$.  We find that the gluon is not changed by
much in the HERA domain $x \gapproxeq 10^{-4}$, compare the
dashed and dotted curves in Fig.~3.  This effect is well known. 
In the power series expansion of the gluon anomalous dimension in
$\alpha_S/\omega$ the coefficients of the 2nd, 3rd and 5th terms are
zero, where $\omega$ is the moment variable.  On the other hand, when
we use pure DGLAP evolution for the quark singlet, as well as
the gluon, the difference is pronounced; compare the dashed and
dot-dashed curves in Fig.~3.  
\begin{figure} 
\centerline{\epsfig{file=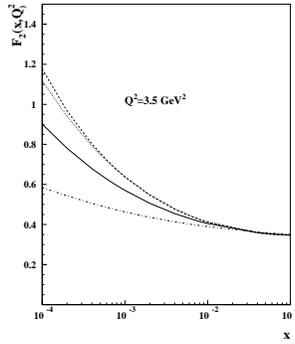,height=5cm,width=5cm}}
\caption{The continuous and dashed curves come from solving
(\ref{eq:b1}) and (\ref{eq:b4}) with and without the kinematic
constraint.  The dotted and dot-dashed curves are obtained using
DGLAP in the gluon sector and in both the gluon and quark sectors
respectively.  The same input is used for all curves.} 
\label{fig.3}
\end{figure}

We would like to conclude that we have a theoretically well-grounded
and consistent formalism which, with the minimum of
non-perturbative input, is able to give a good {\it
perturbative} description of the observed structure of $F_2$. 
Moreover the BFKL/DGLAP components of $F_2$ are decided by
dynamics.  In this way we have made a determination of the {\it
universal} gluon distribution $f (x, k_T^2)$ which can be used,
via the $k_T$ factorization theorem, to predict the behaviour of
other observables at small $x$.  The predictions for $F_2$
(charm) and $F_L$ can be found in \cite{KMS}.

\end{document}